\documentclass[a4paper,fleqn,usenatbib]{mnras}
\pagerange{00--00}
\usepackage{times}
\usepackage{graphicx}
\usepackage{txfonts}
\usepackage{amssymb}
\usepackage{natbib}

\title[K2 photometry and HERMES spectroscopy of $\rho\,$Leo]{K2 photometry and
  HERMES spectroscopy of the blue supergiant $\rho\,$Leo: rotational wind
  modulation and low-frequency waves\thanks{Based on the data gathered with
    NASA's Discovery mission {\it Kepler} and with the {\sc HERMES}
    spectrograph, installed at the Mercator Telescope, operated on the island of
    La Palma by the Flemish Community, at the Spanish Observatorio del Roque de
    los Muchachos of the Instituto de Astrof\'{\i}sica de Canarias and supported
    by the Fund for Scientific Research of Flanders (FWO), Belgium, the Research
    Council of KU\,Leuven, Belgium, the Fonds National de la Recherche
    Scientific (F.R.S.--FNRS), Belgium, the Royal Observatory of Belgium, the
    Observatoire de Gen\`eve, Switzerland and the Th\"uringer Landessternwarte
    Tautenburg, Germany.}}
\author[C.\ Aerts et al.]{C.\ Aerts$^{1,2}$\thanks{E-mail: conny.aerts@kuleuven.be},
D.\ M.\ Bowman$^1$, 
S.\ S\'{\i}mon-D\'{\i}az$^{3,4}$,  
B.\ Buysschaert$^{1,5}$,
C.\ Johnston$^1$,\and
E.\ Moravveji$^1$,
P.\ G.\ Beck$^{3,4}$,
P.\ De Cat$^6$,
S.\ Triana$^{1,6}$,
S.\ Aigrain$^7$,
N.\ Castro$^8$,\and
D.\ Huber$^{9,10,11,12}$,
T.\ R.\ White$^{12}$\\
  $^{1}$Instituut voor Sterrenkunde, KU\,Leuven, Celestijnenlaan 200D, 3001
  Leuven, Belgium\\
  $^{2}$Department of Astrophysics/IMAPP, Radboud University Nijmegen, 6500 GL
  Nijmegen, The Netherlands\\
$^3$ Instituto de Astrof\'{\i}sica de Canarias, 38200, La Laguna, Tenerife, Spain\\
$^4$ Departamento de Astrof\'{\i}sica, Universidad de La Laguna, 38205, La Laguna,
Tenerife, Spain\\
 $^5$ LESIA, Observatoire de Paris, PSL Research University, CNRS, Sorbonne
  Universit\'es, UPMC Univ.\ Paris 06, Univ.\ Paris Diderot, Sorbonne Paris
  Cit\'e, France\\
$^6$ Royal Observatory of Belgium, Ringlaan 3, 1180 Brussels, Belgium\\
$^7$ Oxford Astrophysics, University of Oxford, Denys Wilkinson Building, Keble
Rd, Oxford OX1 3RH, UK\\
$^8$ Astronomy Department, University of Michigan, 311 West Hall, 1085 S.\ University
Ave., Ann Arbor, MI 48109-1107, USA\\
$^9$ Institute for Astronomy, University of Hawaii, 2680 Woodlawn Drive,
Honolulu, HI 96822, USA\\
$^{10}$ Sydney Institute for Astronomy (SIfA), School of Physics, University of
Sydney, NSW 2006, Australia\\
$^{11}$ SETI Institute, 189 Bernardo Avenue, Mountain View, CA 94043, USA\\
$^{12}$ Stellar Astrophysics Centre, Department of Physics and Astronomy, Aarhus
University, Ny Munkegade 120, DK-8000 Aarhus C, Denmark}

\begin{document}

\date{Accepted ?; Received 2017 December ??; in original form ?}


\maketitle

\label{firstpage}

\begin{abstract}
  We present an 80-d long uninterrupted high-cadence K2 light curve of the B1Iab
  supergiant $\rho\,$Leo (HD\,91316), deduced with the method of halo
  photometry.  This light curve reveals a dominant frequency of
  $f_{\rm rot}=0.0373$\,d$^{-1}$ and its harmonics. This dominant frequency
  corresponds with a rotation period of 26.8\,d and is subject to amplitude and
  phase modulation.  The K2 photometry additionally reveals multiperiodic
  low-frequency variability ($<1.5\,$d$^{-1}$) and is in full agreement with
  low-cadence high-resolution spectroscopy assembled during 1800 days. The
  spectroscopy reveals {rotational modulation by a dynamic aspherical
    wind} with an amplitude of about 20\,km\,s$^{-1}$ in the H$\alpha$ line, as
  well as photospheric velocity variations of a few km\,s$^{-1}$ at frequencies
  in the range 0.2 to 0.6\,d$^{-1}$ in the Si\,III\,4567\AA\ line. Given the
  large macroturbulence needed to explain the spectral line broadening of the
  star, we interpret the detected photospheric velocity as due to travelling
  super-inertial low-degree large-scale gravity waves with dominant
  tangential amplitudes {and discuss why $\rho$~Leo is an excellent target
    to study how the observed photospheric variability propagates into the
    wind.}
\end{abstract}


\section{Introduction}

Blue supergiants are in the least understood stage of the evolution of massive
stars. Lack of understanding of this stage
is unfortunate, since the successors of these stars play a key role in
the chemical evolution of their host galaxy. The nucleosynthetic yields after
the blue supergiant stage are strongly dependent on the helium core mass at the
onset of hydrogen shell burning and how the material gets mixed in the stellar
interior during the pre-supernova evolution
\citep[e.g.,][]{Heger2000,Langer2012}. It would thus be highly beneficial if
blue supergiant variability could be monitored and exploited in terms of the
interior physical properties, just as it has recently become possible for
evolved low- and intermediate-mass stars from asteroseismology
\citep{Bedding2011,Mosser2014,Aerts2017a}.  For blue supergiants, this requires
uninterrupted high-precision
space photometry covering months to years, but such data sets are scarce.

Long-term ground-based photometry of mmag-level precision \citep[e.g.,][and
references therein]{vanGenderen1989,Lamers1998} and spectroscopy of km\,s$^{-1}$
precision \citep[e.g.,][]{Markova2005,SimonDiaz2010,Martins2015,Kraus2015}
devoted to blue supergiants typically had sparse sampling and led to aliasing
and high noise levels in the Fourier spectra. As a result, the interpretation of
blue supergiant variability from ground-based data remained
limited. Ultra-violet space spectroscopy was found to be more useful in this
respect. Indeed, time-series spectroscopy from the International Ultraviolet
Explorer revealed narrow and discrete absorption components due to rotational
modulation and wind variability in the line profiles of blue supergiants
\citep{Prinja1986,Massa1995}.  Moreover, the low-cadence mmag-precision
Hipparcos photometry led to the detection of coherent gravity-mode oscillations
in some blue supergiants \citep{Lefever2007}.

With the availability of high-cadence $\mu$mag-level precision space photometry,
a new era has begun for the detection and interpretation of blue supergiant
variability. The few earliest data sets revealed large diversity in behaviour
and periodicities, making it clear that the search for optimal asteroseismology
targets among blue supergiants is challenging
\citep{Lefevre2005,Saio2006,Moffat2008, Aerts2010,Moravveji2012,Aerts2013}.  A
step forward was achieved from combined {\it Kepler\/} photometry and long-term
high-resolution spectroscopy of the O9.5Iab star HD\,188209
\citep{Aerts2017b}. These data revealed low-frequency photospheric variability
due to travelling gravity waves, propagating into the wind.  A similar study for
the more evolved B1Ia supergiant HD\,2905 had to rely on low-cadence Hipparcos
photometry and ground-based spectroscopy and revealed similar variability
\citep{SimonDiaz2018}, although the wind behaviour is more dominant for that
star.

In this {paper}, we report the detection of low-frequency variability in
K2 halo photometry of the bright blue supergiant $\rho\,$Leo (HD\,91316, B1Iab,
V=3.87). We also performed HERMES spectroscopy of the star
with the aim to assess its asteroseismic potential. We present the target in
Section\,2, each of the two new data sets in Sections\,3 and 4, and discuss our
interpretation of $\rho\,$Leo's variability in Section\,5.

\section{The target $\rho\,$Leo}

In their sample study of blue supergiants,
\citet{Crowther2006} obtained the following stellar parameters of 
$\rho\,$Leo:
$T_{\rm eff}\simeq 22\,000\,$K, $\log\,g\simeq 2.55$, $M\simeq 18\,M_\odot$,
$R\simeq 37\,R_\odot$, $\dot{M}\simeq 4\times
10^{-6}\,M_\odot\,$yr$^{-1}$. Recently, \citet{Kholtygin2016}  
concluded from a 3.5\,h high-cadence time series of high-precision
high-resolution spectroscopy that the star reveals non-radial oscillations.
Moreover, these authors deduced that the variability signatures due to these
oscillations propagate into the wind of the star.

In their extensive spectroscopy study of OB stars, \citet{SimonDiaz2014} found
$\rho\,$Leo to have $v\sin i=50\,$km\,s$^{-1}$ and a large macroturbulent
line-broadening of 72\,km\,s$^{-1}$. Such large macroturbulence is naturally
explained by multiperiodic { large-scale} tangential 
velocity fields in the photosphere
\citep{Aerts2009}.  The star was included in the MiMeS survey to search for a
surface magnetic field \citep{Wade2016}, but these data led to a null detection
at the level of a few Gauss (Neiner, private communication) and exclude
surface magnetism as an explanation of the macroturbulence.

\section{K2 Halo Photometry}

Following the idea of scattered-light {\it Kepler\/} photometry of the blue
supergiant HD\,188209 in the nominal Field-of-View of the mission
\citep{Aerts2017b}, the technique of halo photometry was developed and is
meanwhile well established for the follow-up K2 mission
\citep{Pope2016,White2017}.  We hence proposed $\rho\,$Leo for K2 halo
photometry during Campaign\,14 of the mission.  The K2 data set of $\rho\,$Leo
consists of long-cadence (29.4\,min) data covering 79.64\,d.

\begin{figure*}
\begin{center}
\rotatebox{0}{\resizebox{17.cm}{!}{\includegraphics{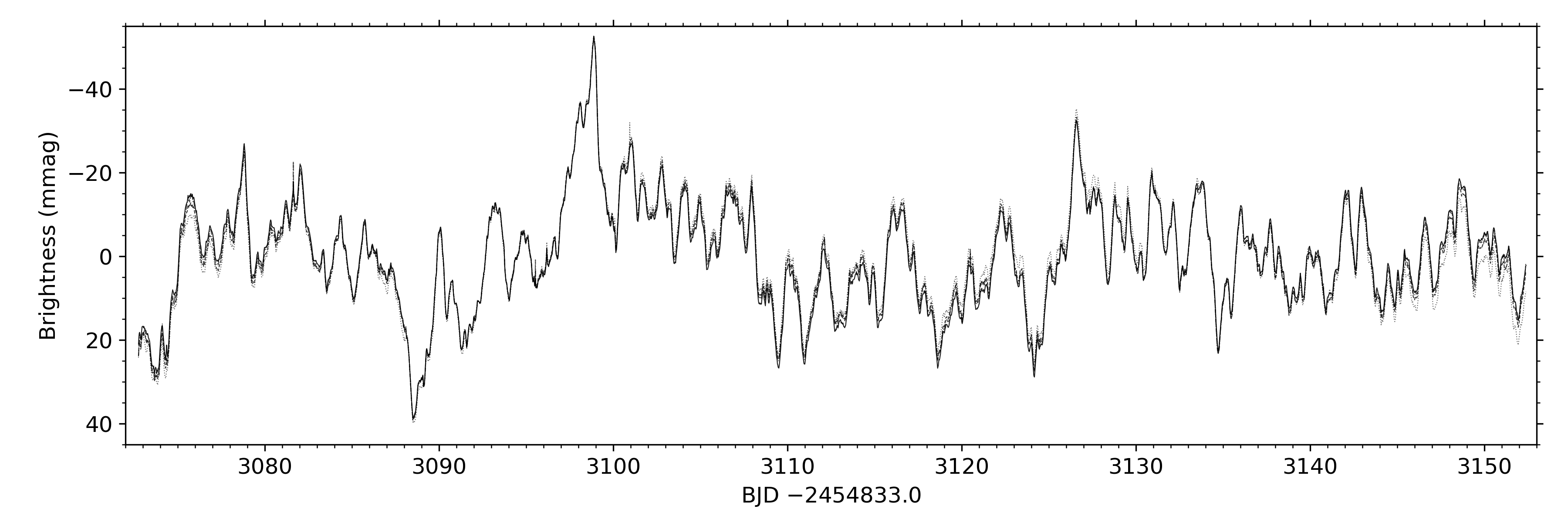}}}
\end{center}
\caption{Four versions of the light curve of $\rho\,$Leo, based on the four
  different halo masks shown in Fig.\,\ref{A1}, are overplotted in
  different line styles. The curves barely show differences, excluding other
  sources of variability in the masks.}
\label{LC}
\end{figure*}

We used the K2 pixel data as provided in MAST\footnote{\tt
  archive.stsci.edu/kepler/data$_{-}$search/search.php} and deduced halo
photometric light curves by adopting four different masks.  The four custom halo
aperture masks were constructed from the stacked images, where the saturated
pixels centered on the star, as well as the very outer regions, were avoided.
We considered the four masks shown in Fig.\,\ref{A1} in Appendix\,A.
Each of the four customised apertures was kept fixed to deduce four versions of
the light curve.  This photometry was subsequently
passed through the {\tt k2sc} software package
\citep{Aigrain2015,Aigrain2016,Aigrain2017} to correct for the pixel drifts and
their related instrumental effects.  An outlier rejection was applied to this
photometry and we finally corrected for a long-term (instrumental) trend by
means of a linear polynomial. The four versions of the light curve, with the K2
flux transformed to brightness expressed in mmag, are overplotted in
Fig.\,\ref{LC}. They are almost indistinguishable from each other. Given the
very different masks, we are confident that the halo photometry reveals the
variability of $\rho\,$Leo itself and is not contaminated by any other
source. The K2 data set has a time base of 79.64\,d, leading to a frequency
resolution (Rayleigh limit) of 0.0126\,d$^{-1}$.

\begin{figure}
\begin{center}
\rotatebox{270}{\resizebox{6.5cm}{!}{\includegraphics{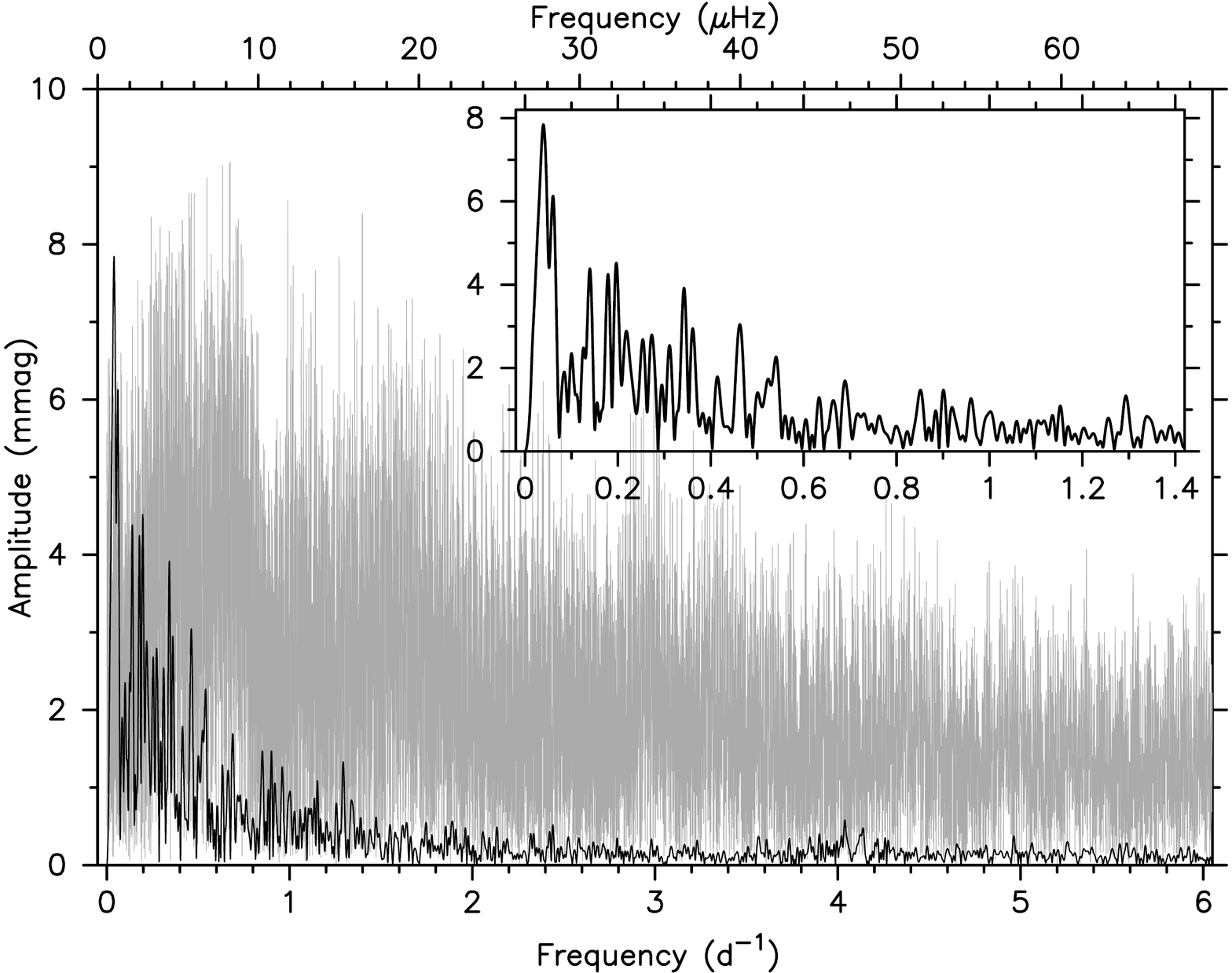}}}
\end{center}
\caption{Scargle amplitude spectra of the K2 halo (thick black lines) and Hipparcos 
(thin grey line) photometry of $\rho\,$Leo.}
\label{FT}
\end{figure}
All frequency analyses of the four light curves shown in Fig.\,\ref{LC} give the
same conclusions; in the rest of the paper we show the results for the full line
in Fig.\,\ref{LC} based on the mask in the upper left panel of 
Fig.\,\ref{A1}.  Scargle amplitude spectra of
the K2 and Hipparcos photometry are shown in Fig.\,\ref{FT}.  
No significant variability due to the star is detected above 
6\,d$^{-1}$.  It can be seen that the
Hipparcos data were of insufficient quality to detect $\rho\,$Leo's photometric
variability, but that the K2 data reveals periodic signal. The dominant
frequency occurs at $f_1=0.0397\pm 0.0002\,$d$^{-1}$ with an amplitude of
$7.94\pm0.28\,$mmag.  The statistical frequency error is far below the
uncertainty due to the low resolving power of 0.0126\,d$^{-1}$.  An unresolved
secondary frequency peak closer than the Rayleigh limit to $f_1$ occurs, in
addition to numerous unresolved low-amplitude frequencies.

\begin{figure*}
\rotatebox{0}{\resizebox{19.cm}{!}{\includegraphics{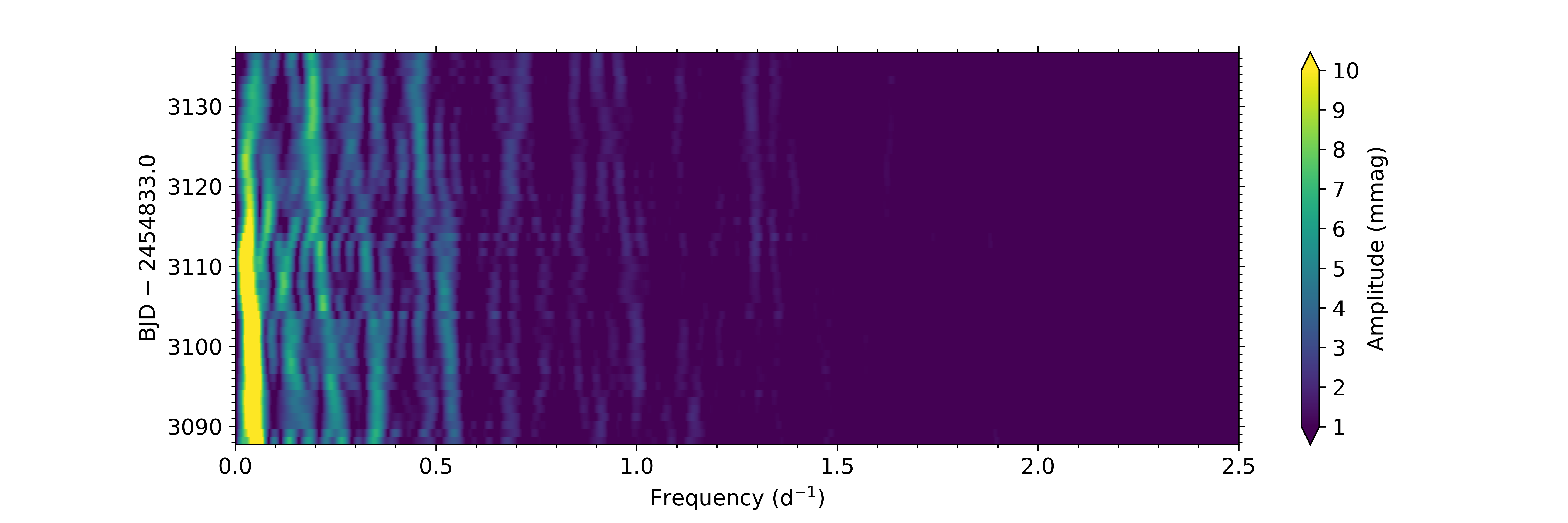}}}
\caption{Short-Time Fourier Transform of the K2 halo light curve computed for a
moving time step of 1\,d and with a window size of 30\,d.} 
\label{STFT}
\end{figure*}
Figure\,\ref{STFT} shows a Short-Term Fourier-Transform (STFT) for the K2 data.
Each of the colour-coded stacked Fourier Transforms is oversampled by a factor
of 10. A moving time step of 1\,d and a window size of 30\,d was used to create
this STFT, but the result is independent of this particular choice. It can be seen that
$f_1$ exhibits amplitude and frequency modulation. This is also the case for
other frequency peaks. Due to the Rayleigh limit, we are
unable to distinguish intrinsic amplitude and frequency modulation of one
feature from multiperiodic beating due to unresolved frequencies.  Hence, any
prewhitening procedure may introduce spurious signal in the residuals, given the
unresolved nature of $f_1$ and its harmonics.  For this reason, we first turn to
sparse but long-term spectroscopy to improve upon the value of $f_1$, before
interpreting the K2 halo photometry further.

\section{HERMES Spectroscopy}

We included $\rho\,$Leo in an ongoing long-term spectroscopic monitoring program
\citep{SimonDiaz2015} 
to assemble multi-instrument data similar to those used in the study of
HD\,188209 \citep{Aerts2017b} and HD\,2905 \citep{SimonDiaz2018}. Here, we limit
our analysis to the currently available 293 spectra taken with the HERMES
spectrograph \citep{Raskin2011} attached to the 1.2m Mercator telescope at La
Palma Observatory, which covers a time base of 1800\,d (Rayleigh limit of
$0.0006\,$d$^{-1}$). Given the mmag variability of $\rho\,$Leo, we have
intensified the ongoing spectroscopic monitoring to reach higher cadence, but
the current HERMES data are sufficient to refine the dominant frequencies
of the star.

\begin{figure}
\begin{center}
\rotatebox{270}{\resizebox{4.cm}{!}{\includegraphics{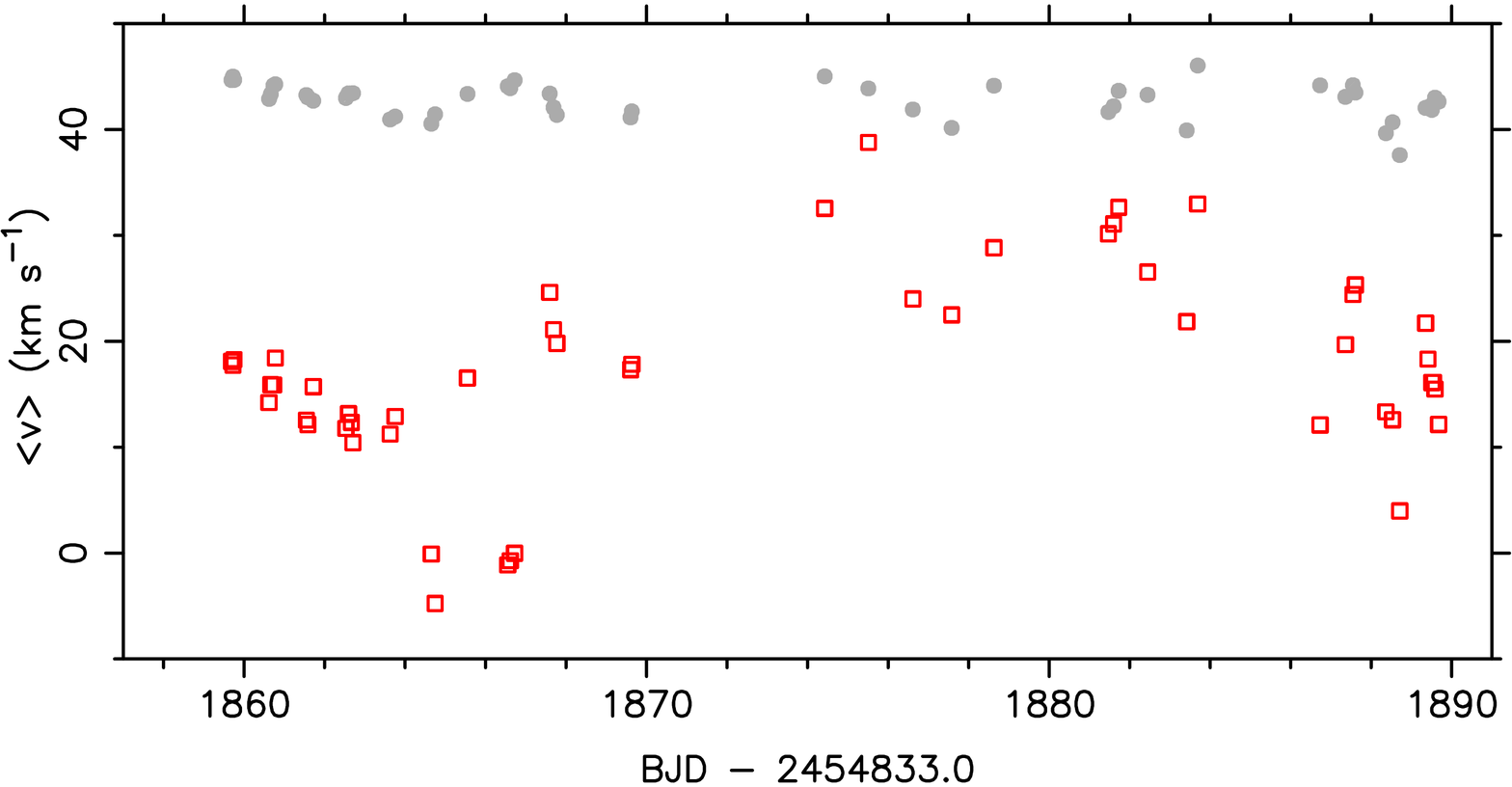}}}\vspace{0.25cm} 
\rotatebox{270}{\resizebox{4.5cm}{!}{\includegraphics{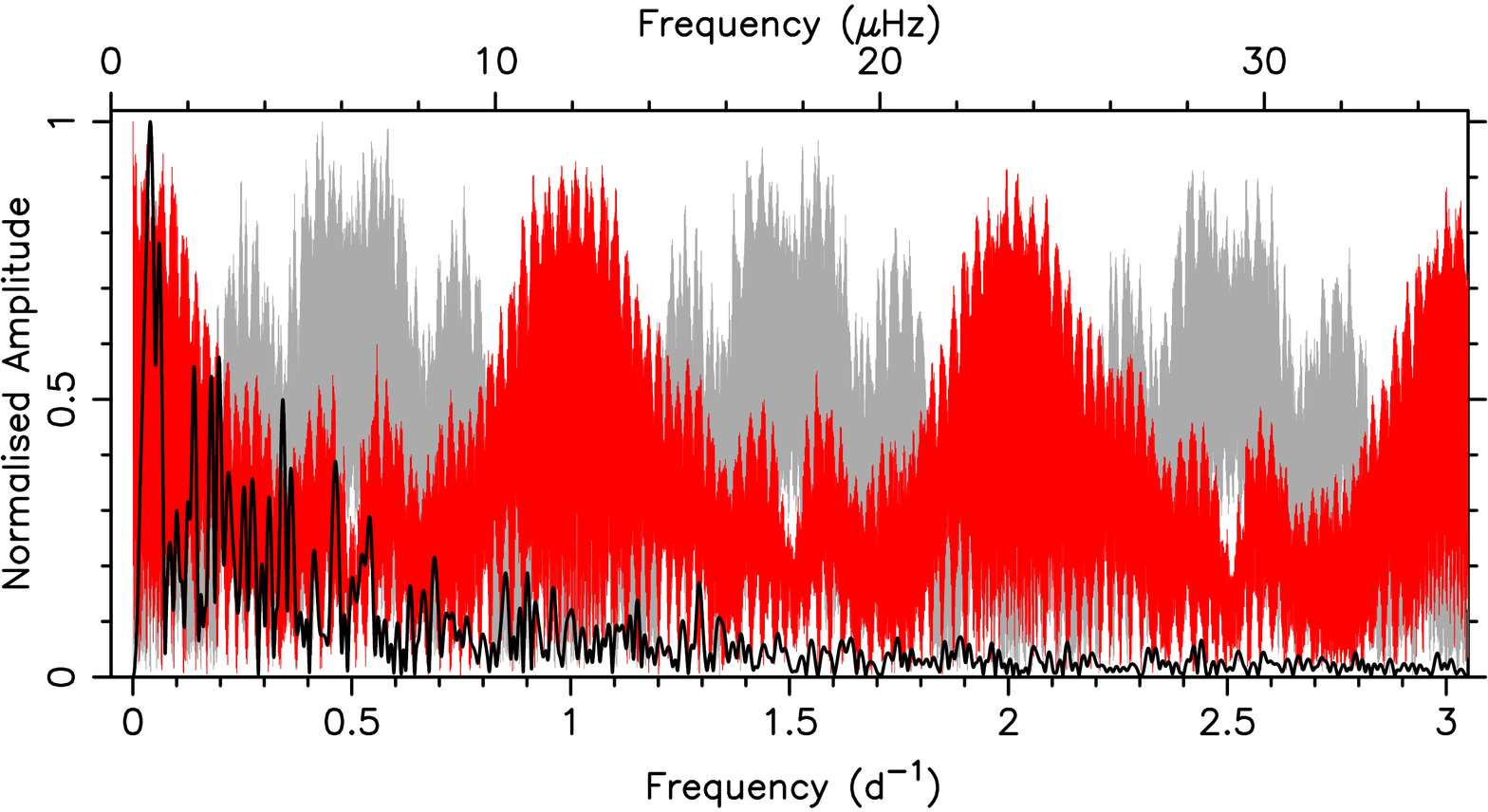}}}
\end{center}
\caption{Top panel: Excerpt of the HERMES radial-velocity data computed as the first moment of the Si\,III\,4567\AA\ line (grey circles) and H$\alpha$ (red squares). Bottom Panel: Scargle spectra normalised to maximal amplitude for the K2 data (black) and of the HERMES $\langle v\rangle$ for the Si\,III\,4567\AA\  line (grey) and H$\alpha$ (red).}
\label{Vrad}
\end{figure}

\begin{figure}
\begin{center}
\rotatebox{270}{\resizebox{4.cm}{!}{\includegraphics{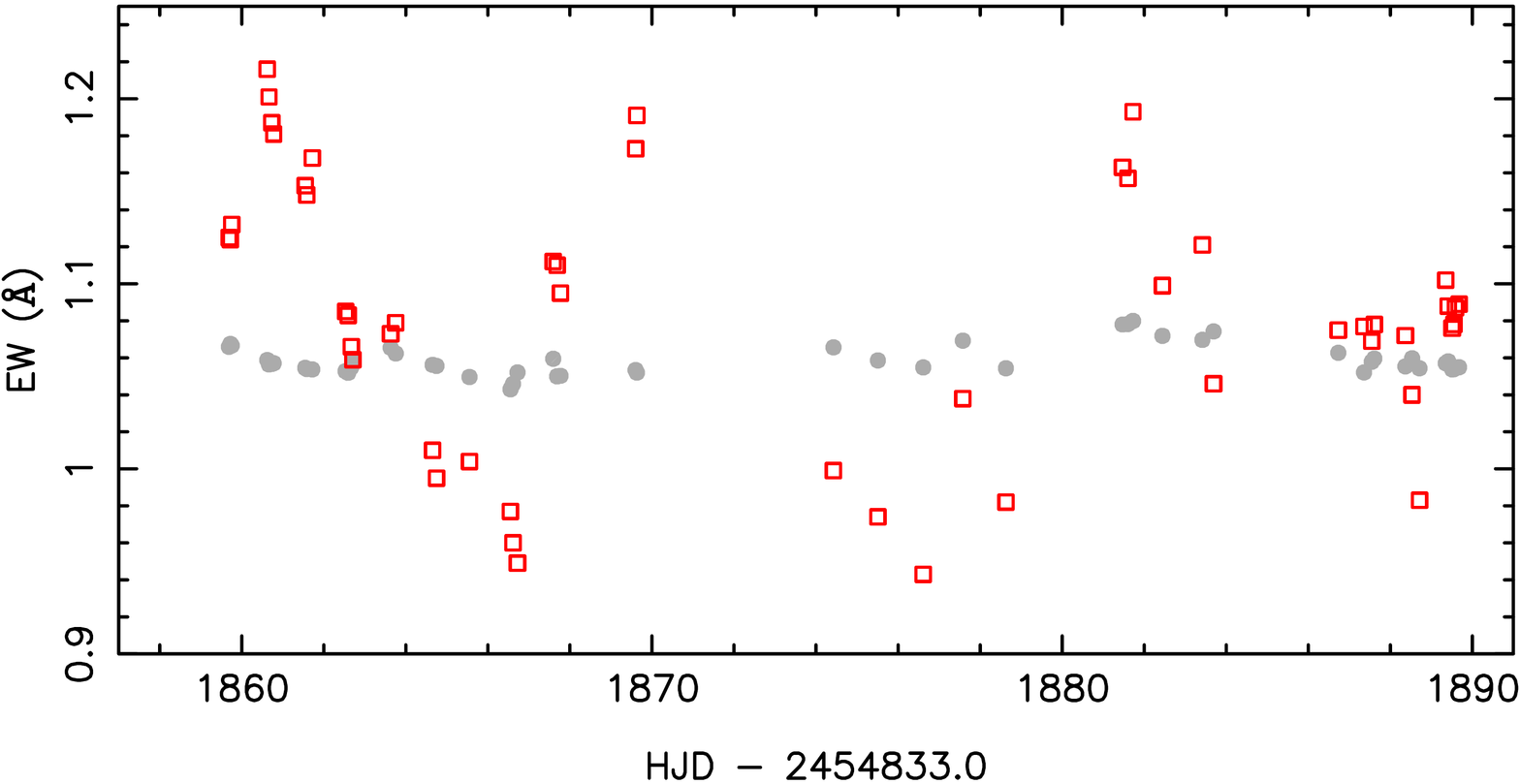}}}\vspace{0.25cm} 
\rotatebox{270}{\resizebox{4.5cm}{!}{\includegraphics{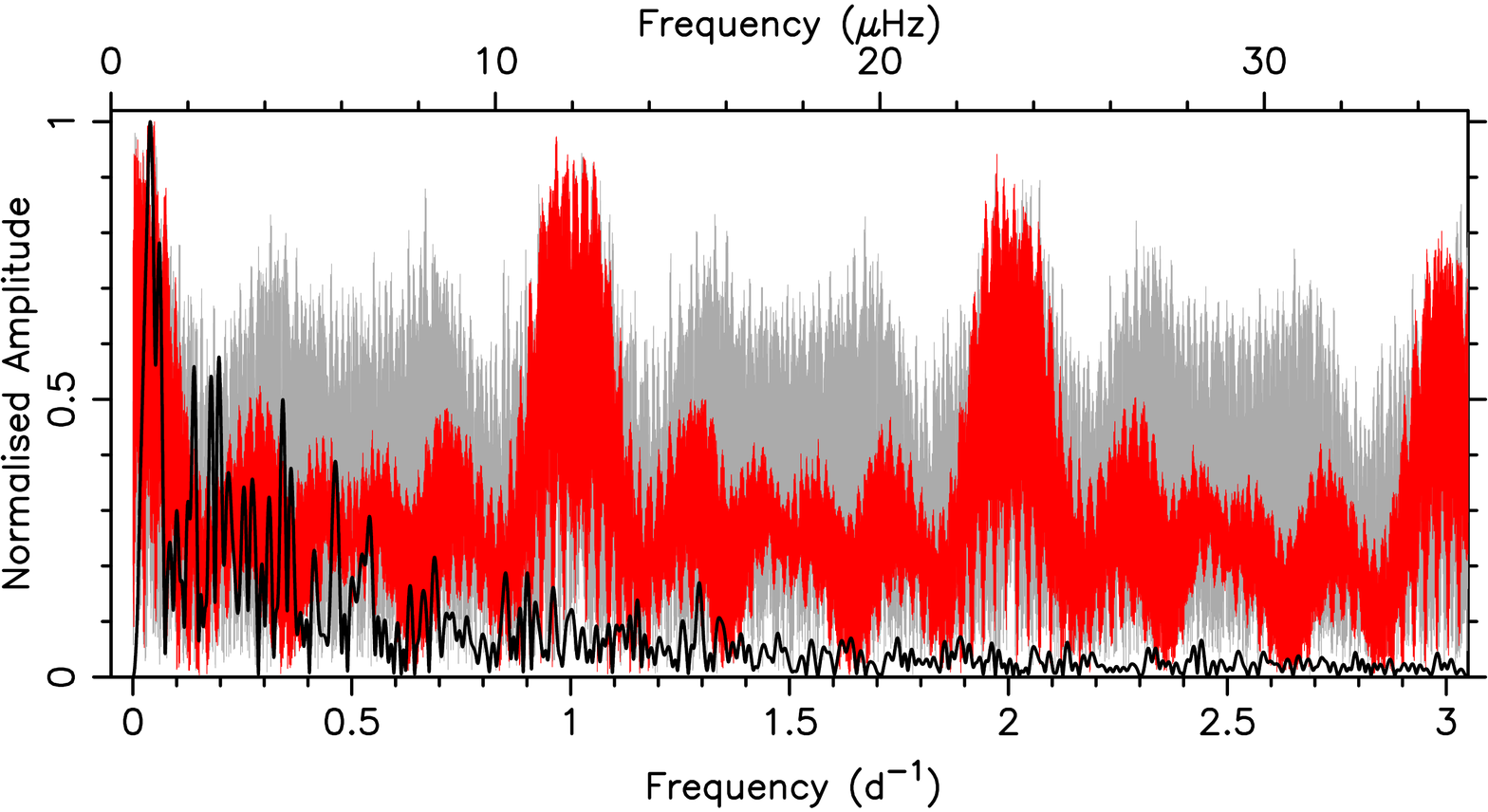}}}
\end{center}
\caption{Top Panel: Excerpt of the HERMES EW data computed as the zeroth moment of the Si\,III\,4567\AA\ line (grey circles, arbitrarily shifted by 0.7\AA\ for visibility purpose) and H$\alpha$ (red squares). Bottom Panel: Scargle spectra normalised to maximal amplitude for the K2 data (black) and of the HERMES EW for the Si\,III\,4567\AA\  line (grey) and H$\alpha$ (red).}
\label{EW}
\end{figure}

\begin{figure}
\begin{center}
\rotatebox{90}{\resizebox{6.cm}{!}{\includegraphics{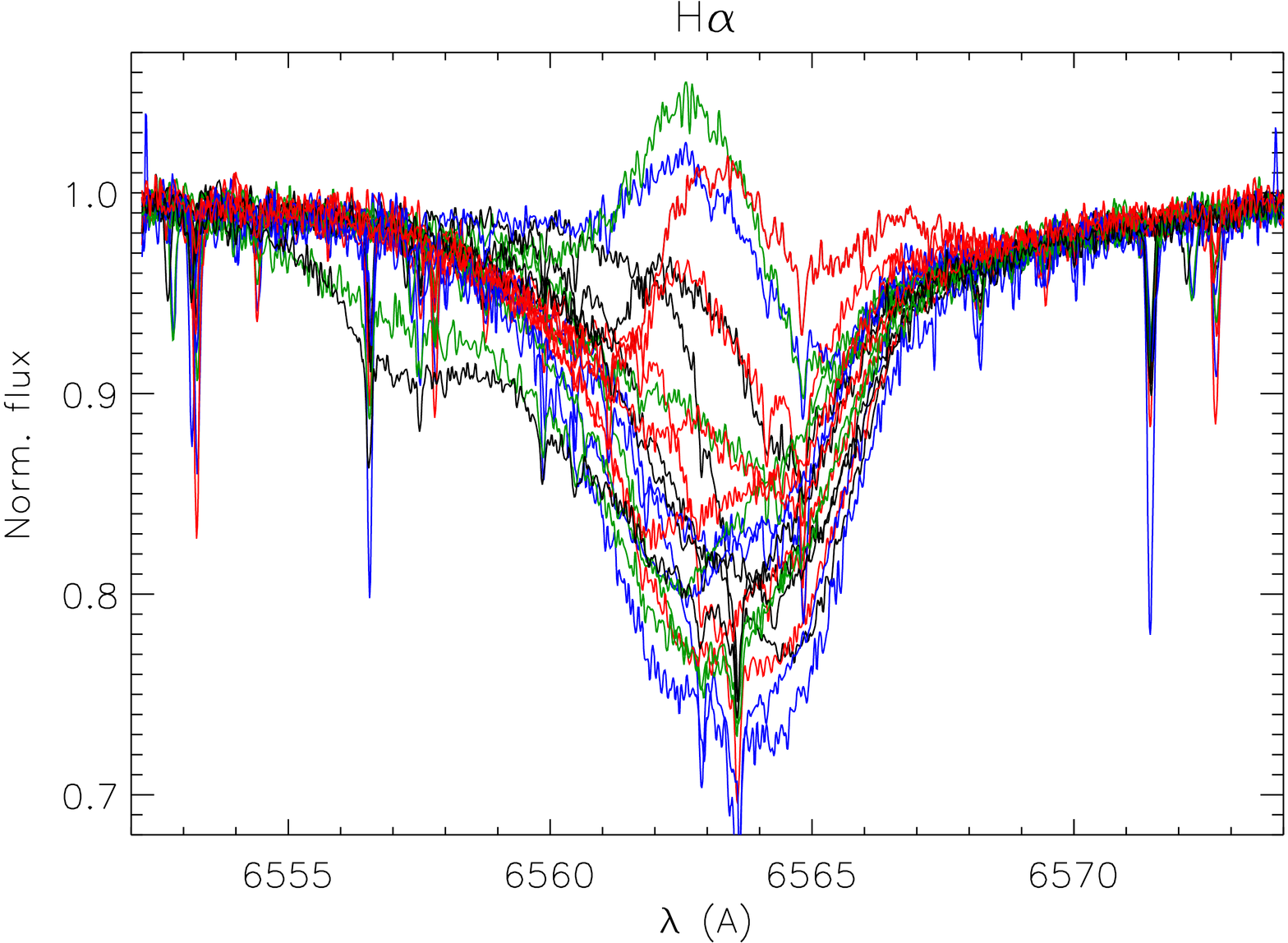}}}
\end{center}
\caption{ {Line profile variability of H$\alpha$ for 20 epochs varying
    between strong line absorption and weak line emission.} }
\label{Halpha}
\end{figure}

{In this initial paper,} we focus on the equivalent width (EW) and radial
velocity $\langle v\rangle$ of the {photospheric line Si\,III\,4567\AA\ and
  the H$\alpha$ wind line to demonstrate that different mechanisms are
  responsible for the variability in these regions of the envelope and
  atmosphere of $\rho$~Leo. We use these two line diagnostics to highlight that
  similar variability to that in the K2 photometry is also present in the
  low-cadence optical spectroscopy.} These two quantities are computed as the
zeroth- and first-order moments of the lines, where EW captures temperature
variations and $\langle v\rangle$ centroid velocity variations
\citep{Aerts1992}.  Figures\,\ref{Vrad} and \ref{EW} show the only ``dense''
part of these otherwise sparsely sampled data and reveal (quasi-)periodicity.
The H$\alpha$ line, formed at the base of the stellar wind, shows complex and
large-amplitude variations with a dominant time scale of $\sim 30\,$d in
$\langle v\rangle$ and about half this period in EW. {To further demonstrate the
  variability in H$\alpha$, we show 20 epochs of spectroscopic observations in
  Fig.\,\ref{Halpha}, which vary between strong absorption and weak
  emission. The time scale and amplitude of this variability differ to that of
  the photospheric Si\,III\,4567\AA\ line, which can also be seen in
  Figs~\ref{Vrad} and \ref{EW}.}

\begin{figure}
\begin{center}
\rotatebox{0}{\resizebox{8.5cm}{!}{\includegraphics{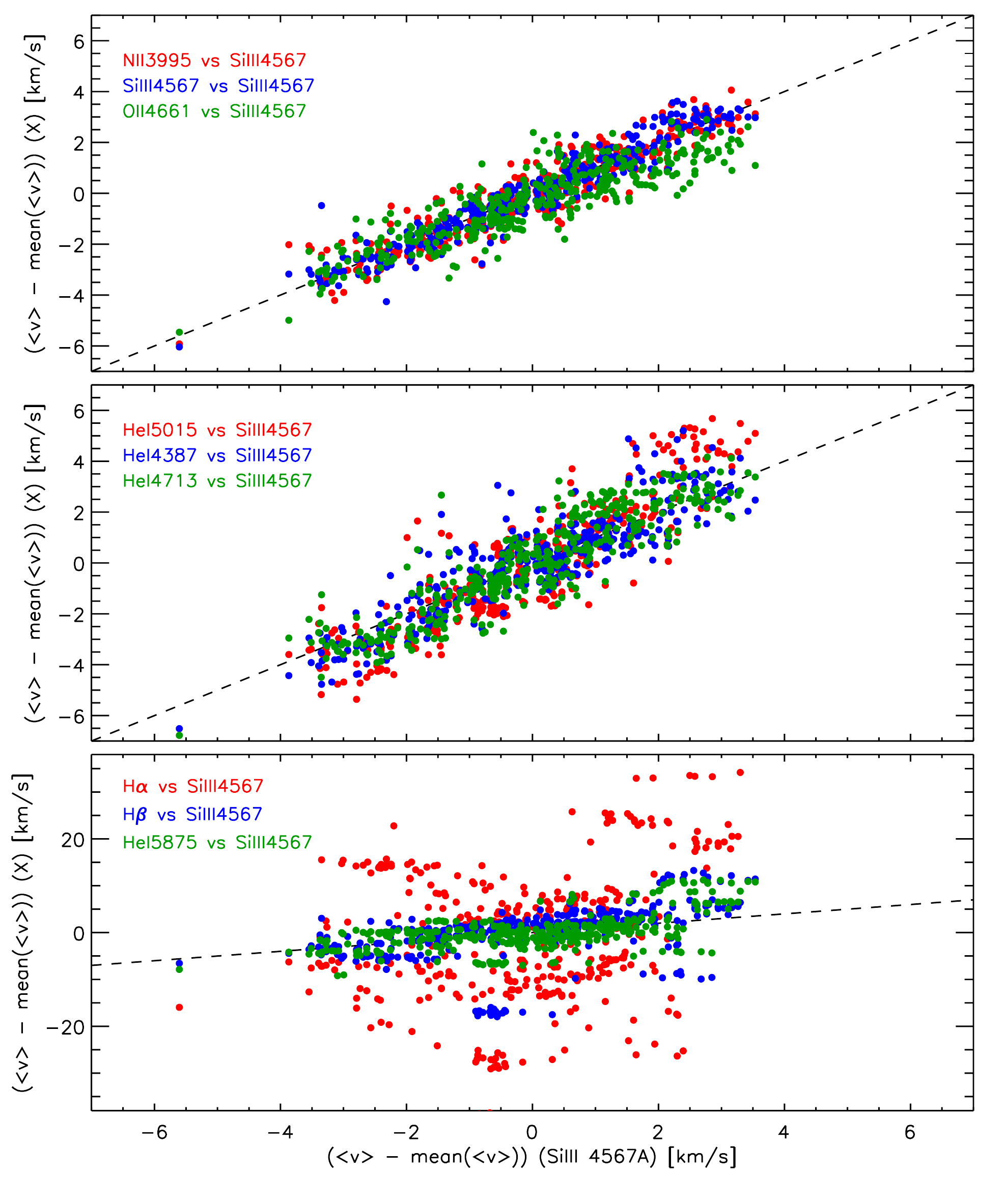}}}
\end{center}
\caption{Correlations in the temporal variability of the first moment (centroid)
  of various diagnostic lines of $\rho\,$Leo.  The Si\,III\,4567\AA\ line is
  used as baseline {as indicated by the axis labels}.  Top and middle panel:
  photospheric lines formed at similar depth than the baseline; bottom panel:
  lines formed at the base of the wind. {The dashed lines in each
    panel represent the bisector.} }
\label{multiple-lines}
\end{figure}

Just as in \citet[][their Fig.\,6]{SimonDiaz2018}, we studied the
  variability of the zeroth and first moments of various lines formed in the
  photosphere in addition to Si\,III\,4567\AA. Even though our spectroscopic
  monitoring of the star is still ongoing, the current HERMES spectroscopy
  already reveals that the moments of various photospheric lines are strongly
  correlated with each other, as illustrated in Fig.\,\ref{multiple-lines}.
  Similar coherent periodicity was found in photospheric lines of the early-B
  supergiant HD\,2905 \citep{SimonDiaz2018}.  Furthermore, we investigated the
  zeroth and first moment variability of other lines partially formed in the
  wind, including H$\beta$ and H$\gamma$, and found that their variability is
  different from the one of the photospheric lines (see also
  Fig.\,\ref{multiple-lines}).  This indicates that the dynamic wind of
  $\rho$\,Leo alters the observed variability in the photosphere, again as in
  HD\,2905 \citep{SimonDiaz2018}. A more detailed analysis using continuing but
  higher-cadence spectroscopy will be provided in a future follow-up paper.  For
  the purposes of the present paper, the photospheric line Si\,III\,4567\AA\ and
  the H$\alpha$ wind line are fully representative and sufficient to probe the
  variability mechanism in the photosphere and wind of $\rho$\,Leo.

We now turn to the Fourier analysis of the EW and $\langle v\rangle$ of the
  Si\,III\,4567\AA\ and the H$\alpha$ lines.  Scargle amplitude spectra of these
four full data sets are heavily affected by daily aliasing but the EW of both
lines are dominated by a frequency close to $f_1$ and $2f_1$, within the
Rayleigh limit of the K2 data. This is also the case for the H$\alpha$
$\langle v\rangle$ data; the Si $\langle v\rangle$ data indicate a dominant
low-amplitude ($\simeq 1.6\,$km\,s$^{-1}$) frequency near
$\simeq0.43\,$d$^{-1}$. In each of the individual data sets, these frequencies
reach low significance, typically between 2 and 4\,$\sigma$.


\begin{figure}
\begin{center}
\rotatebox{270}{\resizebox{6.5cm}{!}{\includegraphics{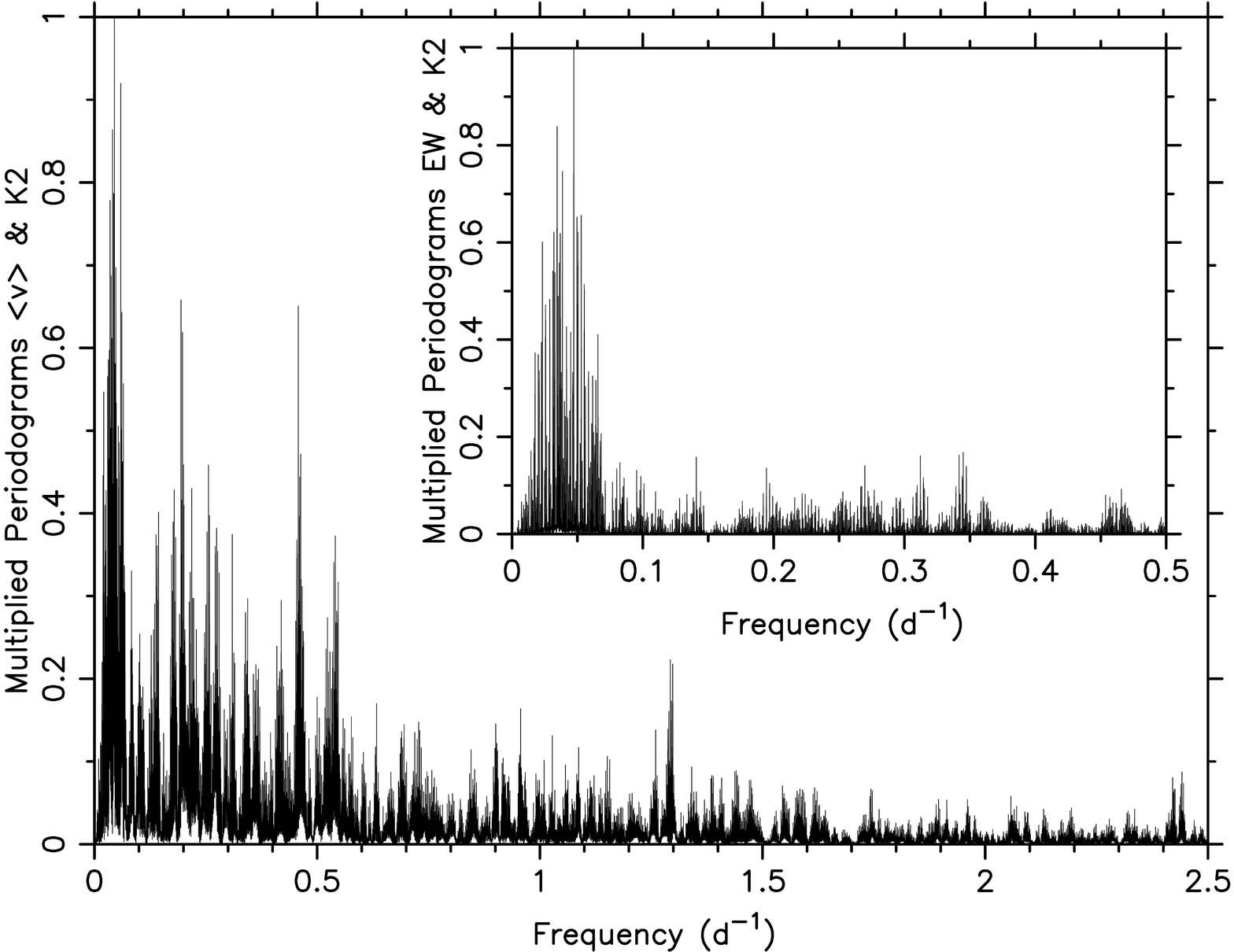}}}
\end{center}
\caption{Multiplied Scargle spectra of the K2 halo photometry of
  $\rho\,$Leo with its $\langle v\rangle$  (and EW: inset) of the H$\alpha$ and
  Si\,III\,4567\AA \ lines of the HERMES spectra.}
\label{multiplied-Scargle}
\end{figure}

\begin{figure}
\begin{center}
\rotatebox{0}{\resizebox{8.5cm}{!}{\includegraphics{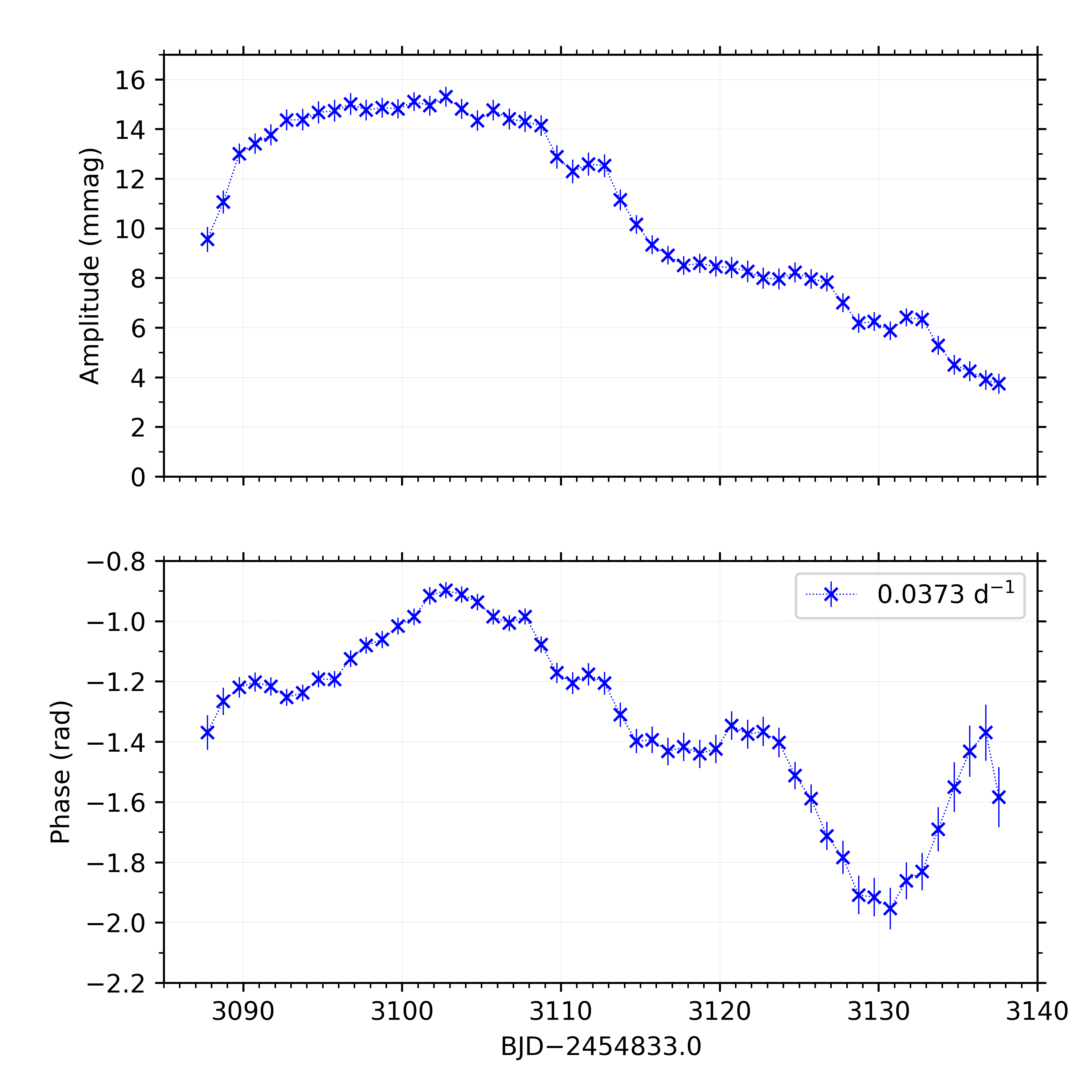}}}
\end{center}
\caption{Amplitude and phase modulation of the dominant frequency across the K2
  light curve.} 
\label{Dom}
\end{figure}

As in \citet{Aerts2017b}, we exploit the full potential of the totally
independent K2 and HERMES data sets by ``merging'' the information in the
Fourier domain. This is done by computing harmonic geometric means of the
normalised dimensionless individual Scargle amplitude spectra, where we give
equal weight to the K2, H$\alpha$, and Si\,III\,4567\AA\ data. In this way, we
combine the high-cadence uninterrupted K2 data with time base of 80\,d with the
low-cadence aliased HERMES photospheric and wind data with time base of
1800\,d. This has the aim to illustrate which frequencies are present in the
multiple datasets, but we refrain from any formal statistical interpretation in terms of
significance.  The result is shown in Fig.\,\ref{multiplied-Scargle} and leads
to a dominant frequency based on EW of $f_{\rm rot}=0.0373\,$d$^{-1}$, well
within the K2 Rayleigh limit of $f_1$. The dominant frequency for
$\langle v\rangle$ is slightly different. This is not surprising as it is
determined by the velocity rather than the temperature variations in the
photosphere. As this frequency gives a much poorer fit to the K2 data, we select
the frequency on the basis of EW and interpret $f_{\rm rot}$ as the best
estimate of the rotation frequency of the star (with corresponding rotation
period of 26.8\,d).  Harmonics of $f_{\rm rot}$ occur in both the
$\langle v\rangle$ and EW multiplied Scargle amplitude spectra. The latter
confirm the findings based on the K2 data alone and revealed by
Fig.\,\ref{STFT}: the dominant K2 frequency is due to cyclic variabillity caused
by rotational modulation {by an aspherical stellar wind} 
and reveals fine structure due to a variable amplitude
and frequency. This is illustrated quantitatively in Fig.\,\ref{Dom}, where we
show the amplitude and phase of $f_{\rm rot}$ computed from the K2 data
following the methodology of \citet{Bowman2016}. The figure represents the
amplitude and phase of $f_{\rm rot}$ by tracking them at fixed frequency using
30-d time bins with a 1-d moving time step. Values of amplitude and phase were
optimised using linear least-squares fits in each time bin and plotted against
time. The zero-point of the time-scale for the phases is BJD 2454943.0. Results
are qualitatively similar for other combinations of bin duration and overlap,
but Fig.\,\ref{Dom} represents the best compromise between frequency and
temporal resolution in the STFT.

The $\langle v\rangle$ multiplied amplitude spectrum in
Fig.\,\ref{multiplied-Scargle} reveals two more frequencies that stand out
beyond the fine structure of the rotational frequency: $f_2=0.1956\,$d$^{-1}$
and $f_3=0.4588\,$d$^{-1}$. These hardly occur in the EW spectrum nor in the
individual Scargle spectrum of the H$\alpha$ $\langle v\rangle$.  These two
frequencies are therefore interpreted as due to velocity variations in the
photosphere. Both of them also occur in the K2 data, at low amplitude.  This
suggests a different astrophysical cause compared to $f_{\rm rot}$ and we
interpret these frequencies, along with several others in
Fig.\,\ref{multiplied-Scargle} that peak above 0.2, in Fig.\,\ref{STFT}, as due
to gravity waves { --- whether this photospheric variability is caused by
  coherent gravity-mode oscillations or convectively-driven travelling gravity 
waves},
or a combination thereof, cannot be deduced from the K2 data due to the too
limited time base of 80\,d. { The frequency spectra derived from numerical
  simulations of gravity waves in massive stars have a similar morphology to
  that shown in Fig.\,\ref{multiplied-Scargle}
  \citep{Rogers2013,Aerts2015,Aerts2017b,SimonDiaz2018}, which supports this
  interpretation as the cause of the photospheric variability in
  $\rho$\,Leo. Furthermore, $\rho\,$Leo lies within the parameter space of the
  instability strip on the Hertzsprung--Russell diagram, in which high-degree
  coherent gravity-mode oscillations are predicted \citep[Star ``C'' in Fig.\,3
  of][]{Godart2017}. Gravity waves have dominant tangential amplitudes, which we
  observe for $\rho$\,Leo in spectroscopy of the photospheric lines, 
supporting this interpretation. For either the case of travelling
  gravity waves or coherent gravity modes, our detection of the photospheric
  variability using both high-quality K2 space photometry combined with
  high-resolution spectroscopy is in full agreement with the similar detection
  by \citet{Kholtygin2016}.}


\section{Interpretation and Discussion}

We unravelled the dominant causes of the variability of the blue supergiant
$\rho\,$Leo from combined 80\,d K2 halo photometry and 1800\,d high-resolution
spectroscopy. All variability occurs at frequencies below 1.5\,d$^{-1}$.  We
find that the dominant variability is caused by rotational modulation at the
base of the {aspherical} stellar wind, with amplitudes of about 8\,mmag in photometry and
20\,km\,s$^{-1}$ in velocity as revealed by H$\alpha$ (Figs\,\ref{LC},
\ref{Vrad}, and \ref{EW}). Variability with amplitudes below 2\,km\,s$^{-1}$ in
velocity and 4\,mmag in brightness are connected with velocity and temperature
variations in the photosphere. These low-amplitude variations are interpreted as
due to gravity waves caused by convective driving or by an opacity mechanism in
the envelope. This interpretation relies on the large macroturbulent velocity
measured in the spectral line wings of the star, which requires tangential
rather than vertical velocity fields.  The velocity variations occur at
frequencies that are an order of magnitude higher than the rotation frequency,
indicating super-inertial large-scale waves of low degree.

{The tangential velocities associated with convectively-driven gravity waves
  in the stellar photosphere of massive stars are of order a tenth of a
  km\,s$^{-1}$ for an individual wave \citep{Rogers2013}, but the combined
  effect of hundreds of gravity waves is similar to the effect of a few
  coherent heat-driven gravity-mode oscillations \citep{Aerts2015}. Thus, both
  travelling gravity waves and coherent gravity modes are feasible explanations
  for the variability observed in the K2 space photometry and high-resolution
  HERMES spectroscopy of $\rho$~Leo, which is similar to the conclusions made
  recently for other O and B supergiants \citep{Aerts2017b,SimonDiaz2018} and
  previously for $\rho$~Leo by \citet{Kholtygin2016}.}

Besides the studies of HD\,188209 (O9.5Iab) and HD\,2905 (B1Ia), combined
long-term space photometry and high-resolution spectroscopy was recently
also assembled
for the fast rotator $\zeta$ Puppis \citep[spectral type
O4I(n)fp,][]{Ramia2017}. The BRITE photometry revealed cyclic variability at the
base of the stellar wind with a rotation period of $1.78\,$d, as well as
stochastic low-amplitude variability assigned to sub-surface convection.  A
quantitative comparison between the frequency spectra caused by
convection/granulation velocities must await 3D convection simulations in the
envelope of OB stars.  2D numerical simulations of gravity waves do lead to the
appropriate horizontal velocity fields that explain macroturbulence
\citep{Aerts2009,Aerts2015}, but this must yet be proven for velocities due to
convection. A key question is whether surface convection can give rise to
large-scale tangential velocity fields in the appropriate frequency range.

Should convective motions cause the detected variability, one also expects some
level of granulation to occur in addition to stochastic variability, as is the
case for acoustic waves excited in the envelope of red giants.  Using the
approximation of granulation scales from \citet{Kallinger2010} for the mass,
radius and T$_{\rm eff}$ estimates for $\rho\,$Leo, and assuming that the solar
values can be scaled, leads to granulation signal at frequencies
$\lesssim 1.5\,$d$^{-1}$.  With the current K2 data set at hand, we cannot
explicitly rule out near-surface convection and granulation as (one of) the
causes of the detected multiperiodic variability of $\rho\,$Leo. However, the
shape of the detected frequency spectrum is more in line with gravity waves, because
we see isolated frequency peaks that resemble a discrete spectrum of eigenvalues
of the star.

{We are currently in the process of obtaining further high-resolution and
  high-cadence 
  spectroscopy of $\rho$\,Leo. In a follow-up
  study, we will carry out a detailed quantitative analysis of} various spectral
lines that are formed at different depths in the photosphere and in the
wind. This will allow {the derivation of the} tangential versus vertical
velocity structure, as well as {how the photospheric variability,
  concluded to be non-radial gravity-modes oscillations by \citet{Kholtygin2016}
  and/or travelling gravity waves in this work, propagates into the wind. The
  wind in $\rho$\,Leo is clearly a dynamic and turbulent environment, as
  indicated by the line profile variability of H$\alpha$ shown in
  Fig.\,\ref{Halpha} varying between strong absorption and emission. Furthermore,
  the variability of other lines formed partially in the wind (i.e., H$\beta$
  and H$\gamma$) vary differently with smaller amplitudes than that of
  H$\alpha$ but larger than those of pure photospheric lines. 
Therefore, $\rho$\,Leo is a promising target to investigate how
  photospheric variability propagates into the wind (see e.g.,
  \citealt{Kholtygin2016}).}

Our study once more illustrates
the power of combined high-cadence uninterrupted space photometry with a time
base of months and ground-based high-resolution spectroscopy covering several
years. Such a combination is necessary for a proper interpretation of hot
supergiant variability and the hunt for optimal asteroseismic targets among blue
supergiants.

\section*{Acknowledgements}
The research leading to these results has received funding from the European
Research Council (ERC) under the European Union’s Horizon 2020 research and
innovation programme (grant agreement N$^\circ$670519: MAMSIE).  S. S-D.\
acknowledges funding by the Spanish Ministry of Economy and Competitiveness
(MINECO) under the grants AYA2010-21697-C05-04, AYA2012-39364-C02-01, and 
Severo
Ochoa SEV-2011-0187.  P.G.B.\ acknowledges support through the MINECO-programme
`Juan de la Cierva Incorporacion' (IJCI-2015-26034).  D.H.\ acknowledges support
by the Australian Research Council's Discovery Projects funding scheme (project
number DE140101364) and support by the National Aeronautics and Space
Administration under Grant NNX14AB92G issued through the {\it Kepler\/}
Participating Scientist Program.  T.R.W.\ acknowledges the support of the Villum
Foundation (research grant 10118).  This paper includes data collected by the K2
mission.  Funding for K2 is provided by the NASA Science Mission directorate.
The authors wish to thank the K2 Guest Observer Office for all their efforts and
support to the scientific community.  The K2 data presented in this paper were
obtained from the Mikulski Archive for Space Telescopes (MAST). Support for MAST
for non-HST data is provided by the NASA Office of Space Science via grant
NNX09AF08G and by other grants and contracts.  We thank the referee for reading
the paper and providing constructive suggestions for improving its content.

\bibliographystyle{mnras}
\bibliography{rhoLeo}

\appendix

\section{Additional figures: 
K2 masks used for the light curve extraction}
\begin{figure*}
\includegraphics[width=0.49\textwidth]{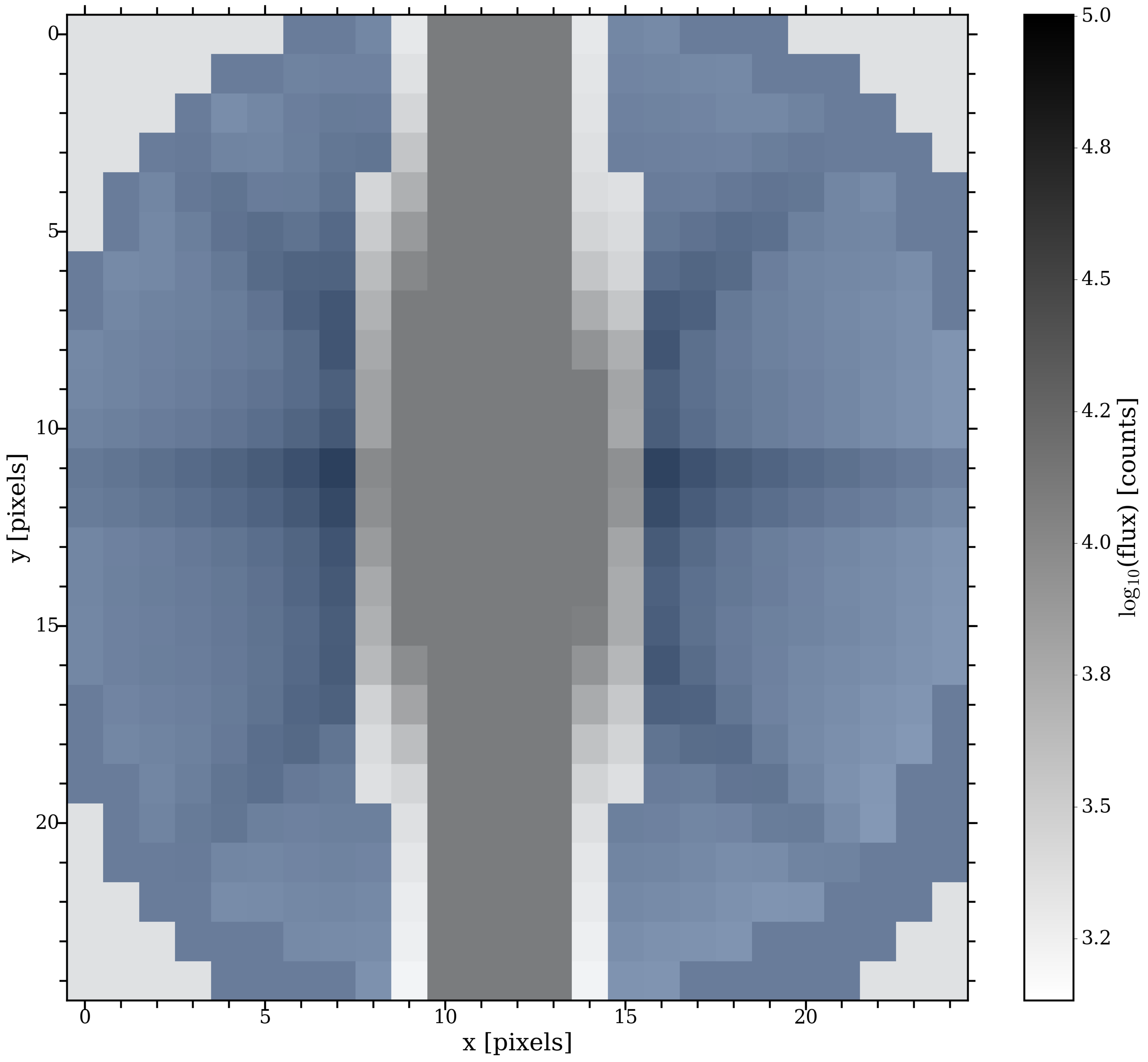}
\includegraphics[width=0.49\textwidth]{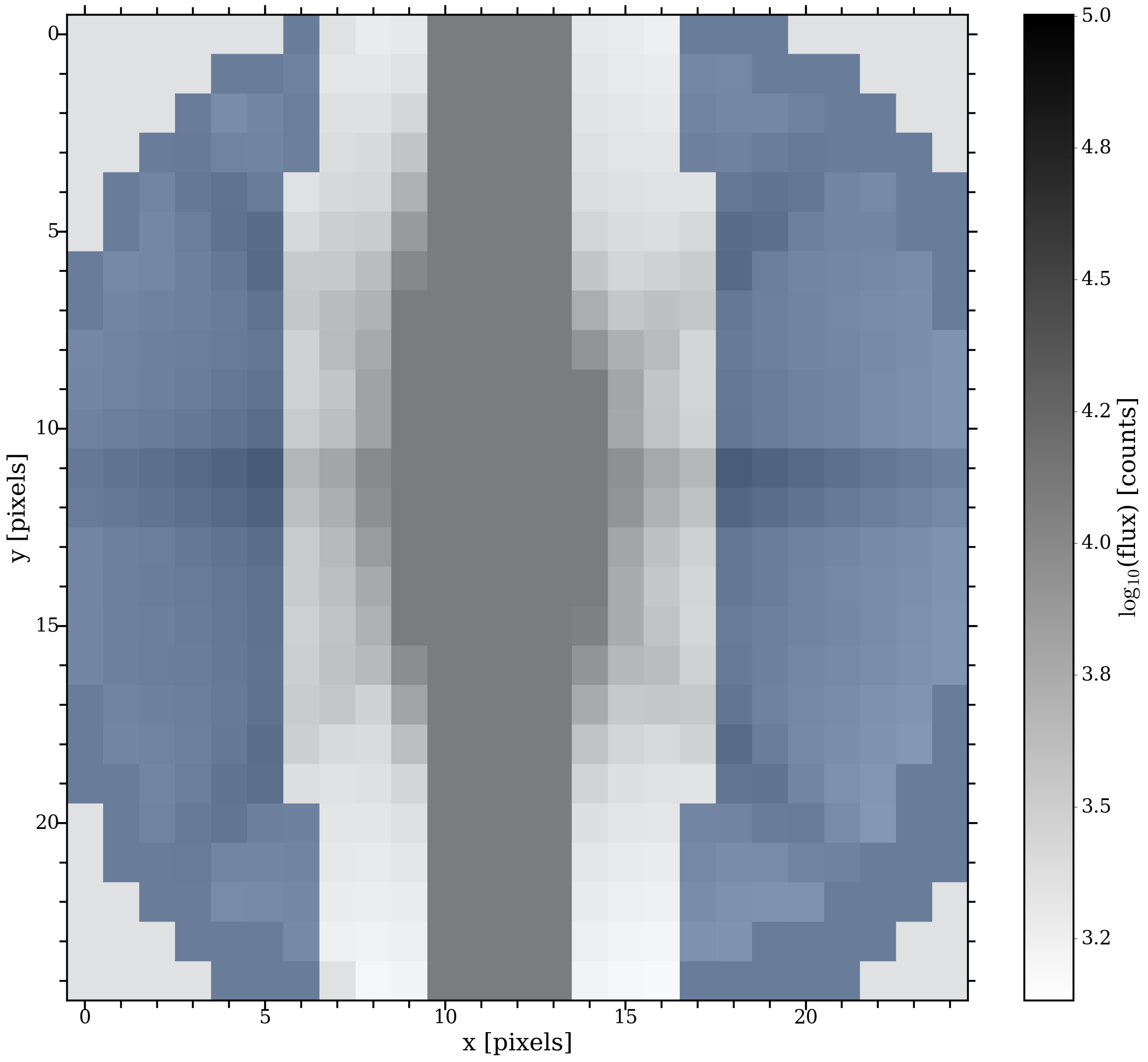}
\includegraphics[width=0.49\textwidth]{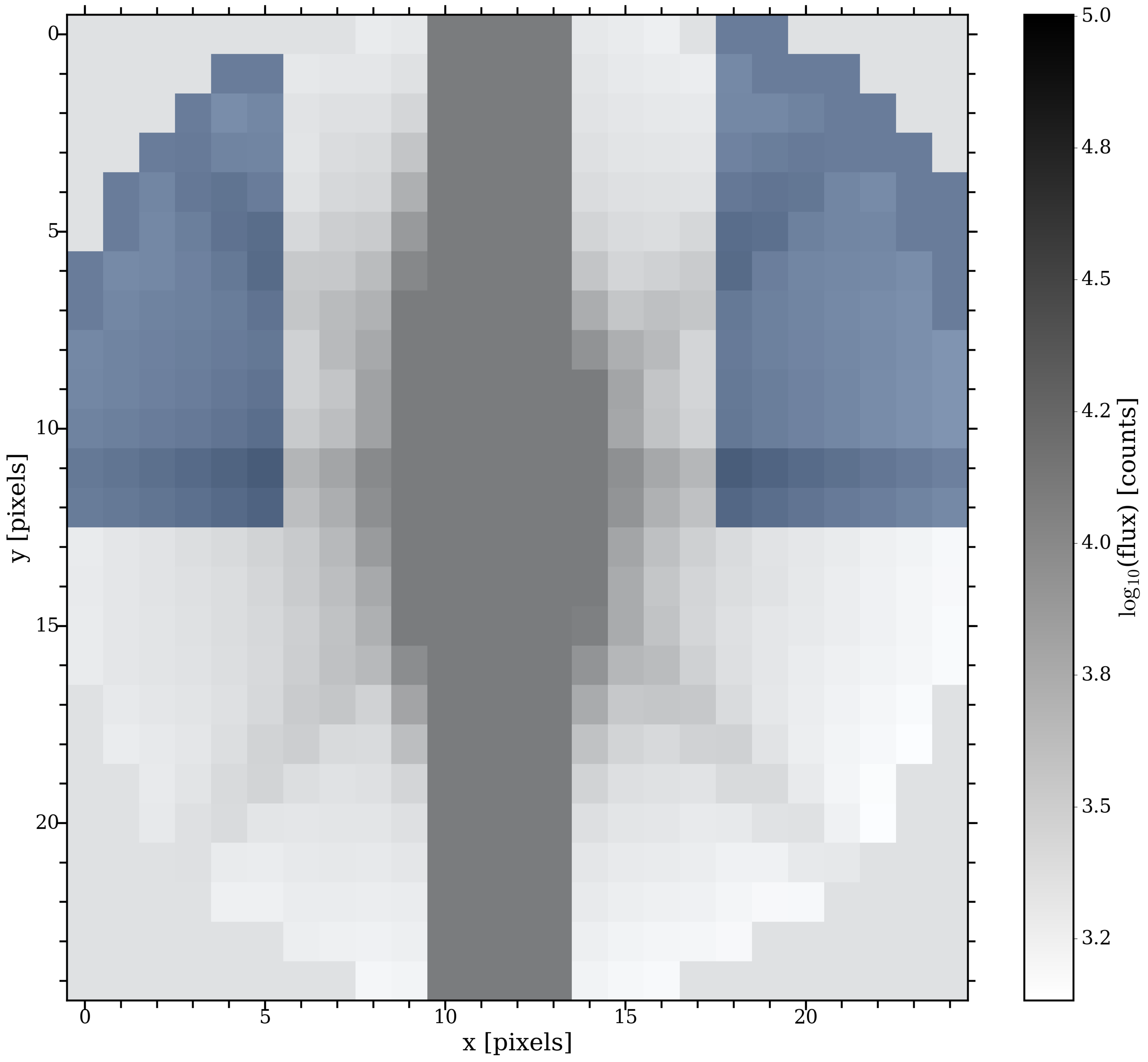}
\includegraphics[width=0.49\textwidth]{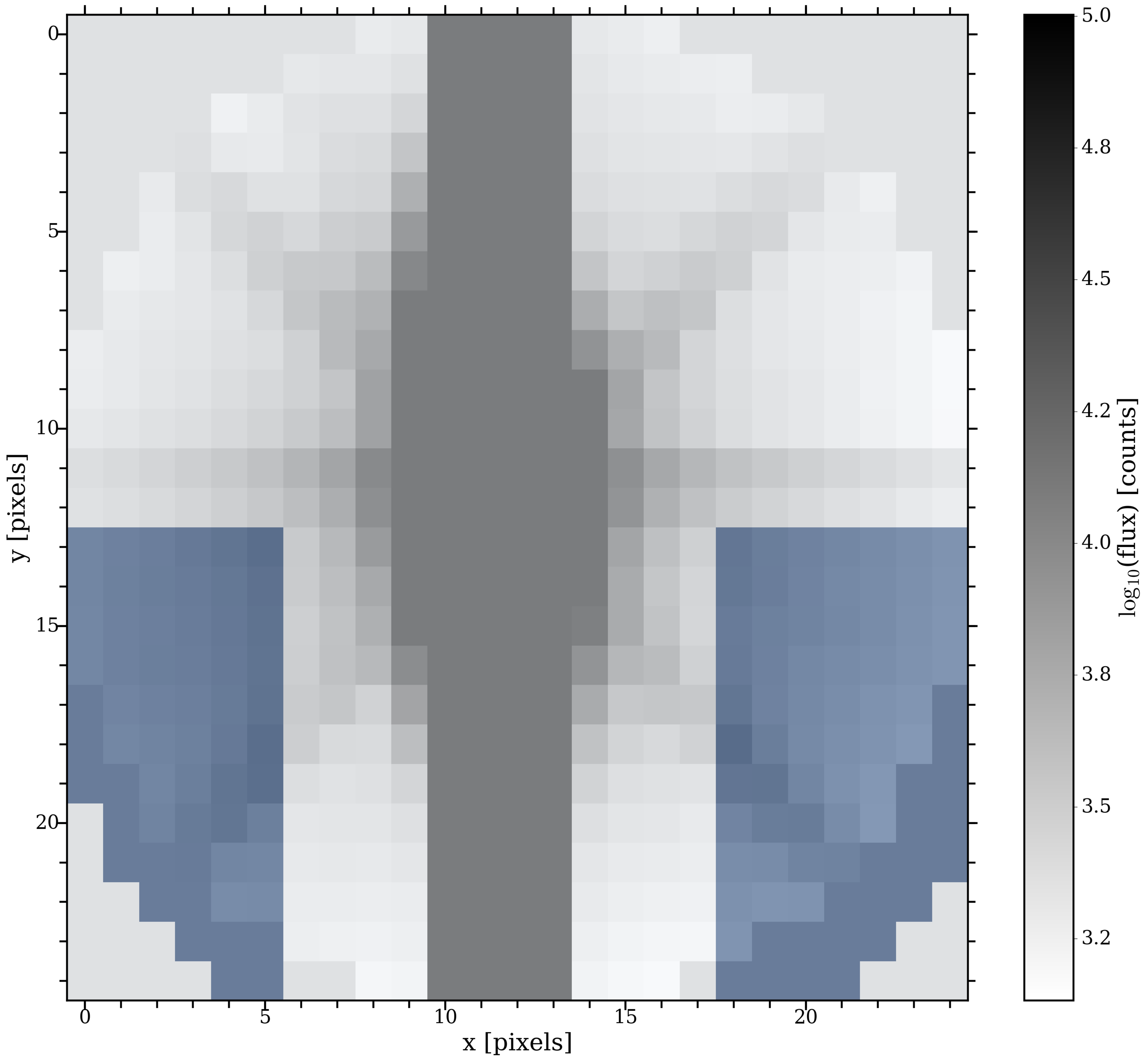}
\caption{The four investigated halo masks used for the light curve extraction of
  K2 photometry of $\rho\,$Leo are shown in blue in each panel, which do not
  include saturated pixels. The chosen mask is shown in the top-left panel with
  the extracted light curve corresponding to the full line in Fig.\,\ref{LC}.}
\label{A1}
\end{figure*}

\label{lastpage}

\end{document}